\documentclass[graybox]{svmult}
\usepackage{mathptmx}       
\usepackage{helvet}         
\usepackage{courier}        
\usepackage{type1cm}        
\usepackage{makeidx}         
\usepackage{graphicx}        
\usepackage{multicol}        
\usepackage[bottom]{footmisc}

\makeindex             

\usepackage{graphicx}
\usepackage{amstext}
\usepackage{amsbsy}
\usepackage{etoolbox}
\include{epsf}
\usepackage{amsfonts} 
\usepackage{authblk} 
\usepackage{sectsty}
\usepackage{amsmath,amssymb,epsfig}
\usepackage{amscd}
\usepackage{mathrsfs}
\usepackage{dsfont}
\usepackage[applemac]{inputenc}
\usepackage[english]{babel}
\usepackage{enumitem} 
\usepackage[]{latexsym}
\usepackage{braket}

\newcommand{\be}{\begin{equation}}
\newcommand{\ee}{\end{equation}}
\def\beqa{\begin{eqnarray}}
\def\eeqa{\end{eqnarray}}
\def\bean{\begin{eqnarray*}}
\def\eean{\end{eqnarray*}}
\def\nn{ \nonumber}

\newcommand{\R}{\mathbb{R}}
\newcommand{\C}{\mathbb{C}}

\newcommand{\tg}{{\tilde{g}}}
\newcommand{\te}{{\tilde{e}}}
\newcommand{\tQ}{{\tilde{Q}}}
\newcommand{\tI}{{\tilde{I}}}
\newcommand{\tJ}{{\tilde{J}}}
\newcommand{\tX}{{\tilde{X}}}

\newcommand\gl{\mathfrak}

\newcommand{\ad}{\hbox{ad}}

\newcommand{\eqn}[1]{(\ref{#1})}
\newcommand{\del}{\partial}
\newcommand{\Tr}[1]{\:{\rm Tr}\,#1}

\renewenvironment{thebibliography}[1]
         {\section*{References}\frenchspacing\small
          \begin{list}{[\arabic{enumi}]}
         {\usecounter{enumi}\parsep=2pt\topsep 0pt
         \settowidth{\labelwidth}{[#1]}
         \leftmargin=\labelwidth\advance\leftmargin\labelsep
         \rightmargin=0pt\itemsep=1pt\sloppy}}{\end{list}}

\begin{document}

\title*{A simple model of double dynamics on Lie groups}

\author{Patrizia Vitale}
\institute{Patrizia Vitale \at Dipartimento di Fisica ``E. Pancini'', Universit\`a di Napoli Federico II
and 
INFN-Sezione di Napoli, 
Complesso Universitario di Monte S. Angelo Edificio 6, via Cintia, 80126 Napoli, Italy. \email{patrizia.vitale@na.infn.it}}
\maketitle
\abstract*{
We study the dynamics of the rigid rotator on the group manifold of $SU(2)$ as an instance of dynamics on Lie groups, together with that of a model whose carrier space is the Borel group $SB(2,\C)$, the Lie Poisson dual of $SU(2)$. We thus introduce a parent action on the Drinfel'd double of the above mentioned groups, which describes the dynamics of  a system with twice as many degrees of freedom as the two starting partners. Through a gauging procedure of its global symmetries both the rigid rotor and the dual model are recovered. 
\\
~
\\
{\it keywords: Generalized Geometry, Double Field Theory, T-Duality, Poisson-Lie symmetry}
}
\abstract{
We study the dynamics of the rigid rotator on the group manifold of $SU(2)$ as an instance of dynamics on Lie groups, together with that of a model whose carrier space is the Borel group $SB(2,\C)$, the Lie Poisson dual of $SU(2)$. We thus introduce a parent action on the Drinfel'd double of the above mentioned groups, which describes the dynamics of  a system with twice as many degrees of freedom as the two starting partners. Through a gauging procedure of its global symmetries both the rigid rotor and the dual model are recovered. 
\\
~
\\
{\it keywords: Generalized Geometry, Double Field Theory, T-Duality, Poisson-Lie symmetry}
}


\section{Introduction}

This paper  is based on a lecture given at the conference in honour of Alberto Ibort, "Classical and Quantum Physics: Geometry, Dynamics and Control" at ICMAT, Madrid,  in March  2018, and it is aimed at discussing within a simple example of dynamics over Lie groups, the isotropic rigid rotator (IRR),  the interplay between concepts such as non-Abelian and Poisson-Lie T-duality \cite{Klim, Klim2}, Generalized and Doubled Geometry \cite{hitchin1, hitchin2, gualtieri:tesi, hulled}, Double Field Theory (DFT) \cite{HZ}, in the mathematical framework of Poisson-Lie groups and Drinfel'd doubles \cite{drinfel'd, semenov}. The main goal being here to convey the general philosophy, many technical details and in deep calculations are left aside and we refer to \cite{MPV18} for an extended presentation of the results. A generalization to field theory is in preparation and will appear hopefully soon \cite{MPV18_2}.

The cotangent space of a $d$ dimensional Lie group, $G$,  $T^* G\sim G\times \R^d$ , while  providing the carrier space for the Hamiltonian dynamics of many systems of physical relevance,  possesses a very interesting structure from the mathematical point of view, it being itself a Lie group, the semi-direct product of the Lie group $G$ with the dual of its Lie algebra, the Abelian Lie algebra $\mathfrak{g}^*\sim\R^d$, thought of as an Abelian vector group.  Free dynamics over the group manifold is described in terms of momenta, $I_i, i=1,...,d$ which are coordinate variables for the fiber $\mathfrak{g}^*$, with  Hamiltonian evolution governed by Kirillov-Souriau-Konstant (KSK) Poisson brackets $\{I_i, I_j\}= c_{ij}^k I_k$ and $c_{ij}^k$ the structure constants of the Lie algebra $\mathfrak{g}$.

The first interesting, although well known,  remark is thus that, as a Lie group $\mathfrak{g}^*$ is Abelian, but the Poisson algebra of linear functions over $\mathfrak{g}^*$, is non-Abelian and isomorphic to the Lie algebra $\mathfrak{g}$. Moreover the latter can be obtained by linearizing the so-called Poisson-Lie structure over the dual group of $G$  which shall be indicated by $G^*$\cite{marmo:articolo1}.

The tangent bundle $TG$ has the structure of a  group, with  its fiber which is   isomorphic to the Lie algebra $\mathfrak{g}$, typically non-Abelian. Fiber coordinates,  namely  the generalized velocities, are components of left or right invariant vector fields $X_a = \dot Q^i_{(a)} \frac{\del}{\del Q^i}, a=1,...,d$  with non-trivial Lie brackets $[X_a, X_b]= c_{ab}^c X_c $. 

Models which exchange tangent and cotangent bundle coordinates over Lie groups, have been widely studied in the context of field theory of sigma models and the duality which is naturally inherited from the relation between the Lie algebra and its dual, is referred to as non-Abelian, or semi-Abelian duality \cite{Klim}. According to the structure which we have outlined,  the dual model in such context relies on a Poisson algebra of momenta which is Abelian.  The IRR system,  which we are going to investigate in the paper, can be regarded as a $0+1$ dimensional analogue of a non-linear sigma model with target space $SU(2)$. 

Once identified the cotangent space of the Lie group $G$  with a semi-direct product of groups, the second interesting observation is that there exists a well defined procedure for deforming the semi-direct product into a fully non-Abelian group, by introducing a non-trivial Lie algebra structure over $\mathfrak{g}^*$, so that the cotangent space $G\ltimes\R^d$ be replaced by $D\simeq G\cdot G^*$ (the latter trivialization being only local) with $G^*$ the Lie group obtained by $\mathfrak{g}^*$ by exponentiation. The group $D$ is the Drinfel'd double of $G$. Since the role of $G$ and   $G^*$  is symmetric, $D$ can be dually regarded as the double of $G^*$ and hence, as a deformation of the cotangent bundle $ T^*G^*$.\footnote{Notice however that, in case $G$ is a compact group, such as $SU(2)$, its  cotangent bundle is  truly diffeomorphic to its Drinfel'd double, while the cotangent bundle of its dual, $SB(2\C)$ for $SU(2)$,  is only a deformation of the double.} Therefore a dynamical system can be defined with the latter as carrier space of the dynamics. The  new algebra of momenta, being dictated by the  KSK  Poisson brackets $\{\tilde I^i,\tilde I^j\}= f^{ij}_k \tilde I^k $ ($f^{ij}_k$ the structure constants of $G^*$), can be retrieved by linearizing the Poisson-Lie bracket over $G$ \cite{MPV18}. Therefore, the IRR model and the latter one are dual to each other, in the sense that  they are defined over partner groups in the Drinfel'd double $D$, with linearized Poisson-Lie brackets of momenta. This is a kind of  Poisson-Lie duality, although the name  has already been used in the literature in the context of sigma models, implying a more stringent relation, which entails the metrics of the two models as well \cite{Klim2}.  Completely integrable  systems and their relation to double Lie groups have been studied in \cite{MI98}. 

A natural step further is to consider a parent action over the Drinfel'd double $D$ with twice as many degrees of freedom as the two actions over $G$, $G^*$ respectively. Such a doubling will generate a Double Field Theory when considering the 1+1-dimensional analogue of the IRR model \cite{MPV18_2}. The generalized action encodes the global symmetries of both models and can be reduced to either of them by means of an appropriate  gauging of some of the symmetries. The generalized momenta ${\bf P_I}$ being sections of the cotangent bundle $T^*D$ can be described in terms of the momenta of the cotangent spaces of the dually related groups $G, G^*$:  ${\bf P_I}\simeq (I_i, \tilde I^i)$, where the position of indices stresses the dual geometric meaning of the two. Because the cotangent fibre of one group can be identified with the tangent fiber of the other and vice-versa,  the connection  with Generalized Geometry is thus natural.

The latter was first introduced by N. J. Hitchin in Ref. \cite{hitchin1}. In a nutshell, it  consists in replacing the tangent bundle $T$ of a manifold $M$ with  $T\oplus T^*$, a bundle with the same base space $M$ but  fibers given by the direct sum of tangent and cotangent spaces. Moreover,  the Lie bracket on the sections of $T$, which are vector fields, is replaced by  the Courant bracket which involves vector fields and one-forms. This is in turn related to  Double Field Theory (DFT)  \cite{HZ}.

DFT was introduced to realize the symmetry of the dynamics of string theory under T-duality transformations   as a manifest symmetry of the string action.  In order  to achieve this goal, the degrees of freedom  of the target space, represented by the coordinate fields   $x^i(\sigma,\tau)$,   $i = 1, \dots, d$ have to be  {\em doubled} with respect to the original.   Therefore,  in the framework of string theory,  the doubling takes place in the $d$-dimensional target space $M$ of the non-linear sigma model underlying the string action, by introducing  new fields $\tilde x_i (\sigma,\tau)$, which are dual to $x^i(\sigma,\tau)$, with $i = 1, \dots, d$.   This is in perfect agreement with the approach of   Generalized Geometry,  by identifying $x^i, \tilde x_i $ with sections of a generalized bundle $E\oplus E^*$ over the world sheet of the string. Thus,  it is only when the target space of the string becomes  the configuration space of an  {\it effective field theory}, that the doubling of fields is reinterpreted as a doubling of the configuration space. A formulation of the string action with manifest T-duality invariance was first proposed by \cite{Tseytlin, Duff}. A corresponding doubling of the  space-time degrees of freedom  in the low-energy effective action first occurred in the pioneering of work of Siegel \cite{siegel}. 

The interest for DFT is  naturally not limited to string theory but  it 
is relevant in the broad area of field theory when one deals with duality symmetries of the dynamics which are not manifest at the level of the action. For an early paper see for example  \cite{siegel84}, which addresses the problem of constructing manifestly Lorentz invariant actions for self dual gauge fields by means of the introduction of auxiliary fields in the Lagrangian. 


To summarize, doubling can occur at  different stages:
\begin{itemize} 
\item at  the level of fields on a given configuration space, as in GG;
\item at the level of configuration space coordinates, with fields $\phi$ depending on twice the initial configuration space variables,  $\phi= \phi(x^i, \tilde x_i)$, as for the strings effective dynamics. In this case it is more appropriate to talk about  {\em Doubled Geometry} (DG) \cite{Hull, hulled};
\item at the level of both, fields and coordinates as for  example in DFT of the strings effective dynamics, when dealing with    the symmetries of the generalized action.  
\end{itemize}
There is therefore an interplay between GG and  DG on the one hand and  DFT on the other hand  which, within  the framework we have sketched, emerges from the identification of the appropriate carrier space of the dynamics.
 It is then clear that models whose dynamics takes place on   Lie groups  can be very helpful in better understanding the above mentioned relations, because the notion of Drinfel'd doubles and  Poisson-Lie dual groups, together with their symmetries, is well established \cite{drinfel'd, semenov}. The idea of investigating such geometric structures in relation to duality in field theory has already been applied to  sigma models by Klim\v{c}\'ik and \v{S}evera in \cite{Klim, Klim2, lledo}  where the authors first introduced the notion of Poisson-Lie T-duality (also see \cite{sfetsos}, \cite{falceto}).  DFT on group manifolds with its relation with Poisson-Lie symmetries, has been studied in \cite{hassler}.
 
 Since our model describes an example  of  particle dynamics, the most appropriate doubling within those enumerated above is  the doubling of the configuration space. For the same reason, we shall see that the model considered here is too simple to exhibit symmetry under duality transformation, although a generalization to field theory is possible \cite{MPV18_2}.

 The paper is intended to give a short  account of the results contained in \cite{MPV18}. Therefore the structure strictly follows the former, although being much shorter and less technical. In Sect. \ref{rigidrot} we review  the dynamics of the IRR on the group manifold of the group $SU(2)$.  In Sect.  \ref{drinfel'd}  we shortly introduce the  mathematical framework of  Poisson-Lie groups and Drinfel'd doubles. In  Sect. \ref{dualrot} we introduce a dynamical model on the dual group of $SU(2)$, the Borel group $SB(2,\C)$ and study  its dynamics. 
In Sect. \ref{gensec}  we present the parent action on the Drinfel'd double, which has twice as many degrees of freedom as the two partner models. The latter are recovered by appropriately gauging the global symmetries of the former.  Finally, in Sect. \ref{hamform} we introduce the Hamiltonian formalism for the double model and in \ref{canform} we study in detail the full Poisson algebra, together with  the  Hamiltonian vector fields associated with momenta $(I_i,\tI^i)$.  The Poisson brackets of the  generalized momenta $(I_i,\tI^i)$  can be related to Poisson-Lie brackets on the two dual groups. The full Poisson algebra of momenta is isomorphic to the algebra of $SL(2,\C)$, namely a semisimple group, with each set of momenta underlying  a non-Abelian algebra. That is why we refer to the two models as non-Abelian duals giving rise to a kind of Poisson-Lie T-duality. 
Sect. \ref{concl} contains our conclusions.

 \section{The Isotropic Rigid Rotator}\label{rigidrot}
The Isotropic Rigid Rotator (IRR) provides   a classical example of dynamics over  Lie groups (see \cite{marmosaletan} as a reference text for the subject).  The tangent [cotangent] bundle of the group $SU(2)$ can be chosen as  carrier space for the dynamics in the Lagrangian [Hamiltonian] formulation. 
A suitable action for the system  is thus the following
\be
S_0= \int_\R L _0~dt=-\frac{1}{4} \int_\R \Tr ( \phi^{*}(g^{-1} d g) \wedge *_H \phi^*(g^{-1} dg)) =-\frac{1}{4}\int_\R  \Tr (g^{-1}{\dot g})^2  dt \label{lag}
\ee
where  $\phi :t\in \R\rightarrow SU(2)$ is the group-valued dynamical variable, while $\phi^*$ is the induced pull-back map.  $
g^{-1} d g = 2 i \alpha^k e_k $ is the Maurer-Cartan left-invariant  one-form,  $e_k=\sigma_k/2 $ are the $SU(2)$ generators with $\sigma_k$ the Pauli matrices, $\alpha^k$ are the basic left-invariant one-forms, $*_H$ denotes the Hodge  
star operator on the source space $\R$, such that $*_H dt = 1$,  and $\Tr$ the trace over the Lie algebra. Moreover,   
\be g^{-1}{\dot g}\equiv \phi^*(g^{-1}{ d g}) (\Gamma)
\ee
 with $\Gamma=d/dt$ .  The model can be regarded as a $(0+1)$-dimensional field theory which is group-valued. 

From now on, with an abuse of notation, we shall trade $g^{-1} d g$ for its pull-back, and the dynamical variable $\phi$ for $g$, as it is customary in  the  dynamics of fields over Lie groups. The group manifold can be parametrized with  $\R^4$ coordinates  as $g= 2(y^0 e_0 
+i  y^i {e_i})$, with $(y^0)^2+ \sum_i (y^i)^2=1$, $e_0={\mathbb I}/2$.  
By observing that 
\be
g^{-1}  \dot g  =  2i (y^0 \dot y^i-y^i \dot y^0+  {\epsilon^{i}}_{jk} {y^j \dot y^k })e_i    \label{qdot}
\ee
we define the left generalized velocities $\dot Q^i$ as 
\be
\dot Q^i  \equiv
 (y^0\dot y^i-y^i \dot y^0+  {\epsilon^{i}}_{jk} y^j \dot y^k)  \label{genvel}
\ee
so that $g^{-1}  \dot g=2 i \dot{Q}^{i} e_{i}$. 
$(Q^i, \dot Q^i)$ $i=1, \dots ,3$ are therefore  tangent bundle coordinates, with $Q^i$ implicitly defined. From  right-invariant one-forms  one could define right generalized velocities in an analogous way. They  give an alternative  set of coordinates  over the tangent bundle. 

The Lagrangian $L_{0}$ in Eq. \eqn{lag} can be rewritten as:
\be
L_0=\frac{1}{2} \dot Q^i \dot Q^j \delta_{i j}.  
\ee
In the intrinsic formulation  the Euler-Lagrange equations of motion can be restated as 
\be
 {\sf L}_\Gamma  \theta_{L} -d L_0=0   
\ee
with 
\be
\theta_L= \frac{1}{2}\Tr[ g^{-1} \dot g  \;g^{-1} d g]= \dot Q^i \alpha^j \delta_{i j }    
\ee
the Lagrangian one-form and ${\sf L}_{\Gamma}$  the Lie derivative with respect to $\Gamma$. With a little calculation we obtain 
\be
{\sf L}_\Gamma\dot Q^j \delta_{ji} - \dot Q^p \dot Q^q {\epsilon_{ip}}^k\delta_{qk} = {\sf L}_\Gamma\dot Q^j \delta_{ji}= 0 \label{eqmo}
\ee
because of the rotation invariance of the product and the complete antisymmetry of the structure constants of $SU(2)$.  
%
Cotangent bundle coordinates can be chosen to be $(Q^i, I_i)$ with the $I_i$'s denoting the left momenta $
I_i= \frac{\del  L_0}{\del \dot Q^i}= \delta_{i j} \dot Q^j  $. An alternative set of fiber coordinates is represented by the right momenta, which are defined in terms of the right generalized velocities. 

The Legendre transform from $TSU(2)$ to $T^*SU(2)$ yields  the Hamiltonian function:
\be
H_0=[I_i \dot Q^i -L_{0}]_{\dot Q^i=\delta^{ij} I_j }= \frac{1}{2} \delta^{ij}I_i I_j \label{h0}  \,\,.
\ee
By  introducing  a dual basis $\{{e^i}^*\}$ in the cotangent space, such that $\langle{e^i}^*|e_j\rangle =\delta^i_j$, one can consider the linear combination:
\be
I=i\; I_i {e^i}^*. \label{Iform}
\ee 
 The dynamics of the IRR is thus  obtained from the Hamiltonian \eqn{h0} and   the   following Poisson brackets
\beqa
\{y^i,y^j\}&=&0\label{pp}\\
\{I_i,I_j\}&=& {\epsilon_{ij\;}}^k I_k \label{xx}\\
\{y^i,I_j\}&=&\delta^{i}_j y^0 +{\epsilon^i}_{\; jk}y^k \;\;\;  {\rm or ~ equivalently}\;\;\; \{g, I_j\}= 2 i  g e_j \label{ij}
 \eeqa
 which can be easily derived from the  action \cite{MPV18}.
 %

The fiber coordinates  $I_i$  are   associated to the angular momentum components and the base space coordinates $(y^0, y^i)$ to the orientation of the rotator. The resulting system is rotationally invariant since
$ 
\{I_i, H_0\} = 0. $ 
The Hamilton equations of motion for the system  read
\be 
\dot I_i= 0,\;\;\; g^{-1}\dot g= 2 i I_i \delta^{ij} e_j.  \nonumber
\ee
Thus the angular momentum
$I_i$ is a  constant of motion, while $g$ undergoes a uniform
precession. Since the Lagrangian and the Hamiltonian are invariant under right and left $SU(2)$ action, as well-known right momenta are conserved as well, being the model super-integrable.

 Let us remark here that, while the fibers of the tangent bundle $TSU(2)$ can be identified, as a vector space,  with the Lie algebra of $SU(2)$, $\mathfrak{su}(2)\simeq \R^3$, with $\dot Q^i$ denoting vector fields components, the fibers of the cotangent bundle $T^*SU(2)$ are isomorphic to the dual Lie algebra $\mathfrak{su}(2)^*$. As a vector space this is again $\R^3$, but the $I_i$ 's are now components of one-forms. This remark will be relevant in the next section, when the Abelian structure of $\mathfrak{su}(2)^*$ is deformed.
 
As a group, $T^*SU(2)$ is the semi-direct product of $SU(2)$ and the Abelian group $\mathbb{R}^3$, with the corresponding Lie algebra given by: 
\be \label{JJ}
\left[L_i,L_j\right]  =i {\epsilon_{ij}}^k L_k, ~~~~~ \left[T_i,T_j\right] = 0, ~~~~~\left[L_i,T_j\right] =i {\epsilon_{ij}}^k T_k. 
\ee
Thus, the non-trivial Poisson algebra on the cotangent space  reflects the non-triviality of the latter. 

Before closing the section let us mention  that already in Ref. \cite{marmo:articolo1}  the phase space of the rigid rotator was generalized to the semisimple group $SL(2, C)$,  by replacing the Abelian subgroup $R^3$ of the semi-direct product above, with the Borel   group $SB(2,\C)$, namely, passing to  the  {\it double} Lie group of $SU(2)$.  
In next section we shall review the mathematical  construction of  Drinfel'd double Lie groups and their relation with the structures of Generalized Geometry. Our generalization will be however different form the one considered in \cite{marmo:articolo1}.

\section{Poisson-Lie Groups and the Double Lie Algebra $\mathfrak{sl}(2,\mathbb{C})$}\label{drinfel'd}
  
 { A
Poisson-Lie group \cite{semenov, alex, babelon, wein} is a Lie group equipped with a Poisson structure
which makes  the product $\mu :G\times G \rightarrow G$ a Poisson map if
$G \times G$ is equipped  with the product Poisson
structure. Linearization of the Poisson structure at the unit $e$ of
$G$ provides   a Lie algebra structure over the dual algebra  ${\gl g}^*=T^*_{e}(G)$ by the
relation
\begin{equation} 
\label{Liedual}
[d\xi_{1}(e),d\xi_{2}(e)]^{*}=d\{\xi_{1},\xi_{2}\}(e)
\end{equation}
with $\xi_i\in C^\infty(G)$. 
The compatibility condition between the Poisson and Lie structures  of $G$ yields the  relation:
\begin{equation} 
\label{comp} 
\left< [X,Y],[v,w]^*\right>+\left<\ad_v^*X,\ad_Y^*w\right>
-\left<\ad_w^*X,\ad_Y^*v\right>
-\left<\ad_v^*Y,\ad_X^*w\right>+\left<\ad_w^*Y,\ad_X^*v\right>=0\,
\end{equation} 
with $v,w\in\mathfrak{g}^*, X,Y\in \mathfrak{g}$ and $ \ad_X^*, \ad_v^*$ the coadjoint actions of the Lie algebras $\mathfrak{g}, \mathfrak{g}^*$ on each other.
This allows one to define a Lie bracket in ${\gl g}\oplus {\gl g}^*$ through the formula:
\begin{equation} 
\label{Liesuma}
[X+\xi,Y+\zeta]=[X,Y]+[\xi,\zeta]^{*}-ad^{*}_{X}\zeta + ad^{*}_{Y}\xi
+ ad^{*}_{\zeta}X - ad^{*}_{\xi}Y \,\, .
\end{equation}
The symmetry between
${\gl g}$ and ${\gl g}^*$ in~(\ref{comp}) implies that one has also a
Poisson-Lie group $G^*$ with Lie algebra $({\gl g}^*,[\ ,\ ]^*)$ and a
Poisson structure whose linearization at $e\in G^*$ gives the bracket $[\ ,\
]$. $G^*$ is the dual Poisson-Lie group of $G$. 
The Lie group $D$, associated to the Lie algebra ${\gl d}= {\gl g}\bowtie {\gl g}^*$
is the Drinfel'd double group of $G$ (or $G^*$, being the construction symmetric).
\footnote{We denote with the symbol $\bowtie$ the Lie algebra structure of $\mathfrak{d}$ as a sum of two non-Abelian, non commuting subalgebras, each one of them acting on its dual.}   

There is a dual  algebraic approach to the picture above, mainly due to Drinfel'd \cite{drinfel'd}, which starts from a deformation of the semi-direct sum $\mathfrak{g}~\dot\oplus ~\R^n$, with $\R^n \simeq\mathfrak{g}^*$, into a fully non-Abelian Lie algebra, which coincides with $\mathfrak{d}$. Let us review the  construction  for the group $SU(2)$,  whose Drinfel'd double is  the group $SL(2,\C)$ \cite{drinfel'd}. } 


The complex Lie algebra $\mathfrak{sl}(2)$ is completely defined by the Lie brackets of its generators:
\begin{equation}
[t_3,t_1]=2t_1; \quad [t_3,t_2]=-2t_2; \quad [t_1,t_2]=t_3;  
\end{equation}
with
\begin{equation}
t_1=
\begin{pmatrix}
0 & 1 \\
0 & 0
\end{pmatrix}
; \quad t_2=
\begin{pmatrix}
0 & 0 \\
1 & 0
\end{pmatrix}
; \quad t_3=
\begin{pmatrix}
1 & 0 \\
0 & -1
\end{pmatrix}
.
\label{gensl}
\end{equation}

The real algebra $\mathfrak{sl}(2,C)$  is obtained by considering  the complex linear combinations 
\begin{equation} \label{b1}
e_1  =\frac{1}{2}(t_1+t_2)=\frac{\sigma_1}{2}, \;  \;\; e_2  =\frac{i}{2}(t_2-t_1)=\frac{\sigma_2}{2}, \;\;\;
e_3  =\frac{1}{2}t_3=\frac{ \sigma_3}{2}
\ee
\be
b_i= i e_i   \;\;\; i=1,2,3
\ee
We have indeed  
 \be \label{su}
  [e_i,e_j] = i{\epsilon_{ij}}^k e_k,~~~~ 
  {[}e_i,b_j{]}= i{\epsilon_{ij}}^kb_k ,~~~~
 {[}b_i,b_j{]}=-i{\epsilon_{ij}}^ke_k 
 \ee
 with $\{e_i\},  i=1,2,3$, generating the $\mathfrak{su}(2)$ subalgebra.  
  
In a similar way, one can introduce the combinations:
\begin{equation}
\tilde e^1=it_1;\qquad \tilde e^2=t_1; \qquad \tilde e^3=\frac{i}{2} t_3,
\label{basisdouble}
\end{equation}
which are dual  to  the generators \eqref{b1},  with respect to the scalar product naturally defined on $\mathfrak{sl}(2,\mathbb{C})$ as:
\begin{equation}
\Braket{u,v}=2\, {\rm Im}(\,Tr(uv)\,), \quad \forall u,v \in \mathfrak{sl}(2,\mathbb{C}).
\label{psd}
\end{equation}
We have indeed 
\begin{equation}
\Braket{\tilde e^i, e_j}=2\, {\rm Im}(\,Tr(\tilde e^i e_j)\,)=\delta^i_j  \label{eupedown}.
\end{equation}
Hence, $\{\tilde e^j\}$ span  the dual vector space $\mathfrak{su}(2)^*$. This is by itself 
 a Lie algebra, the  special Borel subalgebra $\mathfrak{sb}(2,\mathbb{C})$  with the following Lie brackets:
\begin{equation}
[\tilde e^1,\tilde e^2]=0; \qquad [\tilde e^1,\tilde e^3]=- i \tilde e^1; \qquad [\tilde e^2,\tilde e^3]=-i \tilde e^2  
\end{equation}
which   in  compact form read
\begin{equation}
[\tilde e^i, \tilde e^j]= i {f^{ij }}_k \tilde e^k \label{sb}
\ee
with ${f^{ij}}_k=\epsilon^{ij l}\epsilon_{l3k}$.
Moreover 
\begin{equation}
[\tilde e^i,e_j]= i\epsilon^i_{\,jk}\tilde e^k+ i e_k {f^{ki}}_j  .
\label{liemis}
\end{equation}
 For future convenience we also note that:
\be
\tilde e^i \te^j= -\frac{1}{4} \delta^{i3}\delta^{j3}\sigma_0 +\frac{i}{2} {f^{ij}}_k \te^k.\label{ee}
\ee
Since 
\begin{equation}
\Braket{e_i,e_j}=\Braket{\tilde e^i,\tilde e^j}=0
\ee
both $\mathfrak{su}(2)$ and  $\mathfrak{sb}(2,\mathbb{C})$ are maximal isotropic subspaces of $\mathfrak{sl} (2,\mathbb{C})$   with respect to the scalar product \eqref{psd}. We have $\mathfrak{sb}(2,\mathbb{C})= \mathfrak{su}(2)\bowtie \mathfrak{sb}(2,\C)$. 
Therefore,  the Lie algebra $\mathfrak{sl}(2,\mathbb{C})$ can be split into two maximally isotropic dual Lie subalgebras with respect to a bilinear, symmetric, non degenerate form defined on it.  The couple   ($\mathfrak{su}(2)$, $\mathfrak{sb}(2,\C)$), with the dual structure described above, is a Lie bialgebra. Since the role of $\mathfrak{su}(2)$ and its dual algebra can be interchanged,  $(\mathfrak{sb}(2,\C)$,  $\mathfrak{su}(2)$) is a Lie bialgebra as well.  The triple  $(\mathfrak{sl}(2,\mathbb{C}), \mathfrak{su}(2), \mathfrak{sb}(2,\mathbb{C}))$ is called a  {\it Manin triple} \cite{drinfel'd}. 

$D=SL(2,\C)$ is thus the {\em double group} obtained by exponentiating the bialgebra,  endowed with some additional structures such as a Poisson structure on the group manifold compatible with the group structure. The two partner groups, $G=SU(2)$ and $G^*= SB(2,\mathbb{C})$ with suitable Poisson brackets,  are named  {\it dual groups}. Their role can be interchanged, so that they  share the same double group $D$.  


Besides the scalar product \eqn{psd},  there is another non-degenerate, invariant scalar product,  represented by
\begin{equation}
(u,v)=2 Re( \, Tr(u v) \,) \qquad \forall u,v \in \mathfrak{sl}(2,\mathbb{C}). 
\label{sp2}
\end{equation}
In this case, for the basis elements, one gets:
\begin{equation}
(e_i,e_j)=\delta_{ij}, \quad (b_i,b_j)=-\delta_{ij}, \quad (e_i,b_j)=0,
\label{otherscalar}
\end{equation}
giving rise to a metric which is  not positive-definite. With respect to this, new maximal isotropic subspaces  can be defined, which are spanned by: 
\be
f_i^+=\frac{1}{\sqrt 2} (e_i+ b_i) \,\,\, \quad ; \,\,\, \quad f_i^-= \frac{1}{\sqrt 2}(e_i-b_i) \label{newb}.
\ee
satisfying 
\begin{equation}
(f^+_i,f^+_j)= (f^-_i,f^-_j)=\,\,0 \quad ;  \quad (f^+_i,f^-_j)=\delta_{ij} .
\end{equation}
By denoting by $C_+$ and $ C_-$ the two  subspaces spanned by $\{e_i\}$ and $\{b_i\}$ respectively,  one can notice \cite{gualtieri:tesi} that the splitting ${\mathfrak d}= C_{+} \oplus C_{-}$ defines a positive definite metric $\mathcal{H}$ on ${\mathfrak sl(2,\C)}$ via:
\be
\mathcal{H}= (\;,\;)_{C_+}-  (\;,\;)_{C_-} \label{metricG}
\ee
Let us indicate the Riemannian metric with double round brackets. One has then:
\be
((e_i,e_j)) \equiv (e_i,e_j); ~~~~~ ((b_i,b_j)) \equiv -(b_i,b_j);~~~~~ ((e_i,b_j)) \equiv (e_i,b_j)=0  \,. \label{riem}
\ee

Let us introduce the notation:
\begin{equation}
e_I=\begin{pmatrix}e_i\\ e^i \end{pmatrix},
 \qquad  e_i \in \mathfrak{su}(2), \quad  e^i \in \mathfrak{sb}(2,\mathbb{C}).
\label{doubledb}
\end{equation}
Then  the scalar product \eqn{psd} becomes
\begin{equation} \label{Lprod}
\Braket{e_I,e_J}={\cal \eta}_{IJ}= 
\begin{pmatrix}
0 & \delta_i^j \\
\delta_j^i &  0
\end{pmatrix}
. 
\end{equation}
{This symmetric inner product has signature  $(d,d)$ and therefore defines the non-compact orthogonal group  $O(d,d)$, with $d=3$ in this case}.

The Riemannian product \eqn{riem} yields instead:
\beqa
((\tilde e^i, \tilde e^j))&=&\delta^{ip}\delta^{jq} ((b_p + e_l{\epsilon^l}_{p3} ))((b_q+ e_k{\epsilon^k}_{j3} )) \nonumber \\
&=& \delta^{ij}+\epsilon^i_{\;l3} \delta^{lk}  \epsilon^j_{\;k3} \,\,\,\,\, ; \label{sp2ee}\\
(( e_i, \tilde e^j))&=& [(e_i, b_q) + {\epsilon^k}_{q3} (e_i, e_k)] \delta^{jq}={\epsilon_{3i}}^j \,\,\, .\label{mixed}
\eeqa
Hence,  one has:
\begin{equation} \label{Rprod}
((e_I,e_J))={\cal H}_{IJ}= 
\begin{pmatrix}
\delta_{ij} & \delta_{il}\epsilon^{jl3 }\\
\epsilon^{il3} \delta_{lj}\;  \;& \; \delta^{ij}+ \epsilon^{il3} \delta_{lk}\epsilon^{jk3} 
\end{pmatrix}
. 
\end{equation}
This metric satisfies the relation: 
\be {\cal H}^T {\cal \eta} {\cal H}= \eta \label{compatibile}
\ee
indicating that ${\cal H}$ is a pseudo-orthogonal $O(3,3)$ matrix.


 \section{The Dual Model}\label{dualrot}
In this section we introduce a dynamical model on the dual group of $SU(2)$, the Borel group $SB(2,\C)$,  with an action functional that is  formally analogous to \eqn{lag}. 

 As carrier space for the dynamics  in the Lagrangian (respectively Hamiltonian) formulation one can choose  the tangent (respectively cotangent) bundle of the group $SB(2,\C)$. 
A suitable action for the system  is the following:
\be
{\tilde S}_0= \int_\R {\tilde L} _0~dt= - \frac{1}{4} \int_\R {\mathcal Tr}\,[\tilde\phi^*(\tg^{-1} d \tg\wedge *_H\tilde\phi^*( \tg^{-1} d\tg)] = - \frac{1}{4}\int_\R   {\mathcal Tr} \,[(\tg^{-1}{ \dot \tg})(\tg^{-1}{ \dot \tg})] dt \label{dualag}
\ee
with $\tilde \phi : t\in \R\rightarrow SB(2,\C)$, the group-valued dynamical variable, and $\tilde\phi^*$ the pull-back map. Analogously to the IRR case, 
$
\tg^{-1} d \tg= i \beta_k \te^k  $ represents  the Maurer-Cartan left invariant one-form on the group manifold, with $\beta_k$ the left-invariant basic one-forms,   $*_H$ the Hodge  
star operator on the source space $\R$, such that $*_H dt = 1$. Moreover, as previously, in order to adhere to the notation which is commonly adopted in field theory, we shall identify the dynamical variable $\tilde \phi$ with $\tg$ for the remainder of this section. The symbol ${\mathcal Tr} $ stands for  a {\it suitable} scalar product on  the Lie algebra $\mathfrak{sb}(2,\C)$. Indeed, since the algebra is not semi-simple, there is no scalar product which is both non-degenerate and invariant. We choose to work with  the scalar product induced by the Riemannian metric $\mathcal{H}$, which, on the algebra $\mathfrak{sb}(2,\C)$ takes the form  \eqn{sp2ee} which is positive definite and non-degenerate. This is only invariant under left $SB(2,\C)$ action besides being  $SU(2)$ invariant.
Indeed, by observing that the generators $\te^i$ are not Hermitian, \eqn{sp2ee} can be verified to be equivalent to:
\vspace{0.3cm}
\be
((u,v)) \equiv 2{\rm Re}\Tr [(u)^\dag v]  \label{2ndprod}
\ee
 which 
is not invariant under right $SB(2,\C)$ action. 

 Similarly to the IRR case, the model can be regarded as a $(0+1)$-dimensional field theory which is 
 group-valued. 
 
The group manifold can be parametrized with  $\R^4$ coordinates as  $\tg= 2( u_0 \te^0 
+ i  u_i \te^i)$, with $u_0^2- u_3^2=1$ and $\te^0= {\mathbb I}/2$.  
By observing that 
\be
\tg^{-1} \dot \tg=2 i  (u_0\dot u_i-u_i \dot u_0+  {f_{i}}^{\,jk} {u_j \dot u_k })\te^i  \label{tiqdot}
\ee
the Lagrangian in \eqn{dualag} can be rewritten as:
\be
\tilde{{L}}_0= (u_0\dot u_i-u_i \dot u_0+  {f_{i}}^{\,jk} {u_j \dot u_k })(u_0\dot u_r-u_r \dot u_0+  {f_{r}}^{\,pq} {u_p \dot u_q })((\te^i,\te^r)) 
= \dot \tQ_i \dot \tQ_r  h^{ir}   \nonumber
\ee
with 
$
\dot \tQ_i \equiv u_0\dot u_i-u_i \dot u_0+  {f_{i}}^{\,jk} {u_j \dot u_k }  \nonumber
$
the left generalized velocities
and
\be
h^{ir} \equiv (\delta^{i r}+ {\epsilon^i}_{l3}{\epsilon^r}_{s3}\delta^{ls}) \label{hir}
\ee
the metric defined by the scalar product. $(\tQ_i, \dot \tQ_i)$ are therefore  tangent bundle coordinates, with $\tQ_i$ implicitly defined. 

By repeating the analysis already performed for the IRR,  one finds the equations of motion:
\be
{\sf L}_\Gamma(\dot \tQ_j\, i_{\tX^i}\beta_l )h^{jl} - {\sf L}_{\tX^i} \tilde{L}_0 = {\sf L}_\Gamma\dot \tQ_j h^{ji} - \dot \tQ_p \dot \tQ_q {f^{ip}}_k h^{qk} = 0.  \label{LGamma}
\ee
with $\tX^j$ being the left invariant vector fields generating the right action of  $SB(2,\C)$. Differently from the IRR case, the second term in the RHS is not vanishing, because the structure constants are not completely antisymmetric. This is to be expected because the Lagrangian  is not invariant under right action. 

It has to be noticed here that, analogously to the IRR case, one could  define the  right generalized velocities on the fibers starting from right invariant one-forms, but, differently from that case, the right invariant Lagrangian is not equivalent to the left invariant one.  

 The cotangent bundle coordinates are $(\tQ_i, \tI^i)$ with $\tI^i$ the conjugate left momenta 
$
\tI^j= \frac{\del {\tilde{{ L}}}_0}{\del \dot \tQ_j}= \dot \tQ_r (\delta^{j r}+ \epsilon^j_{\,l3}\epsilon^r_{s3}\delta^{ls}).
$
On inverting for the velocities $\dot\tQ_j= \tI^i(\delta_{ij}-\frac{1}{2}\epsilon_i^{\,p3}\epsilon_j^{\,q3}\delta^{pq}),  $ we get   the Hamiltonian function: 
\be
\tilde{H}_0=[\tI^j \dot \tQ_j -\tilde L]_{\dot \tQ=\dot \tQ(\tI)}= \frac{1}{2}\tI^i (h^{-1})_{ij }\tI^j \,\,\, ,\label{h0du}
\ee
with 
\be
 (h^{-1})_{ij } \equiv (\delta_{ij}-\frac{1}{2}\epsilon_i^{\,p3}\epsilon_j^{\,q3}\delta_{pq}).
 \label{hinvij}
 \ee
On introducing  the linear combination:
$
\tilde I= i \tilde I^j {\te_j}^* 
$
with $\langle {e_j}^* |\te^i\rangle=\delta_j^i$ we obtain 
 the first order dynamics by means of the following Poisson brackets:
\beqa
\{u_i,u_j\}&=&0\label{ppdu}\\
\{\tI^i,\tI^j\}&=& {f^{ij\;}}_k\tI^k \label{xxdu}\\
\{u_i,\tI^j\}&=&{\delta_{i}^j u_0 -{f_i}^{\; jk}u_k \;\;\; {\rm or ~ equivalently}\;\;\; \{\tg, \tI^j\}= 2 i  \tg \te^j }\label{ijdu}
 \eeqa
which are derived from the first order formulation of the action functional (see \cite{MPV18} for details). 

Specifically  we get:
 \be
 \dot\tI^j= \{\tI^j,\tilde H\}= f^{jk}_l\tI^l\tI^r h^{-1}_{kr}   
 \ee
 which is  different from  zero, as expected, expressing the non-invariance of the Hamiltonian under right action. 
 Vice-versa, by introducing the right momenta $\tJ^i$ 
 one readily obtains:
 \be
 \dot\tJ^j= \{\tJ^j,\tilde H\}= 0
 \ee
 namely, right momenta are constants of the motion and the Hamiltonian is invariant under left action, as it should. 
 Right momenta are therefore conserved, as for the rigid rotator, while left momenta are not.

 The same remark as at the end of Sect. \ref{rigidrot} applies:  while the fibers of the tangent bundle $TSB(2,\C)$ can be identified, as a vector space,  with the Lie algebra of $SB(2,C)$, $\mathfrak{sb}(2,\C)\simeq \R^3$, with $\dot \tQ_i$ denoting vector fields components, the fibers of the cotangent bundle $T^*SB(2,C)$ are isomorphic to the dual Lie algebra $\mathfrak{sb}(2,\C)^*$. As a vector space this is again $\R^3$, but $\tI^j$ are now components of one-forms. 
 
As a group, $T^*SB(2,\C)$ is the semi-direct product of $SB(2,\C)$ and the Abelian group $\mathbb{R}^3$, with Lie algebra the semi-direct sum represented by
\be
\left[B_i,B_j\right]  =  i {f_{ij}}^k B_k , ~~~~~
\left[S_i,S_j\right] = 0, ~~~~~ 
\left[B_i,S_j\right] = i {f_{ij}}^k S_k. \label{BS}
\ee
Then, as for the IRR, the  non-triviality of the  Poisson algebra over the cotangent bundle  of the group $SB(2,\C)$ reflects the structure of the latter. 

Before closing the section, let us summarize the results. We have introduced a model on the dual group of $SU(2)$, the Borel group $SB(2,\C)$, whose Hamiltonian dynamics is retrieved in terms of Poisson brackets of KSK type.    As we shall see, the Poisson brackets of the momenta $I_i, \tI^i$ are dually related.

\section{The generalized action}\label{gensec}
In the previous sections we have introduced two dynamical models on configuration spaces which are dual Lie groups.   The Poisson algebras for the respective cotangent bundles, $T^*SU(2)$, $T^*SB(2,\C)$, which we restate for convenience in the form:
\beqa
\{g,g\}&=&0,\;\;\;\;\{I_i,I_j\}=  {\epsilon_{ij}}^k I_k , \;\;\;\;\; \{g, I_j\}= 2 i  g e_j \label{sudue}\\
\{\tg,\tg\}&=&0,\;\;\;\;
\{\tI^i,\tI^j\}= {f^{ij}}_k\tI^k ,\;\;\;\;
\{\tg, \tI^j\}= 2 i  \tg \te^j  \,\,\, , \label{sbdue}
 \eeqa
have both the structure of a semi-direct sum which reflects  the semi-direct structure of the Lie algebras $\mathfrak{su}(2)\dot\oplus\R^3$ and $\mathfrak{sb}(2,\C) \dot \oplus \R^3$. 

In order to unify the two models within a generalized action, whose configuration space has {\it double}  dimension with respect to the previous ones,   let us introduce the configuration space  variable  $\Phi:t\in \mathbb{R} \rightarrow \gamma \in SL(2,\mathbb{C})$.  The left invariant one-form on the group manifold is then:
\begin{equation}
\Phi^*(\gamma^{-1} \mathrm{d}\gamma)= \gamma^{-1} \dot \gamma\; dt \equiv  \dot {\bf Q}^I e_I \mathrm{d}t \label{gammagamma}
\ee
with $e_I=(e_i, \tilde e^i)$ the $\mathfrak{sl}(2,\C)$ basis introduced in Eq. \eqn{doubledb} and  $ \dot {\bf Q}^I$, the left generalized velocities. As in the previous sections, in order to adhere to the notation which is commonly adopted in field theory, we shall identify the dynamical variable $\Phi$ with $\gamma$ for the remainder of this section. By defining the decomposition $\dot {\bf Q}^I \equiv( A^i, B_i)$ one has:
\be
\gamma^{-1} \dot \gamma \;dt= (A^i e_i + B_i \tilde e^i) dt  \nonumber
\ee
where, however, both components are tangent bundle coordinates for $SL(2,C)$. 
By using the scalar product \eqn{psd}, the components of the generalized velocity can be explicitly obtained:
\be
A^i= 2{\rm Im} \Tr ( \gamma^{-1}\dot \gamma \tilde e^i); \;\;\; B_i= 2{\rm Im} \Tr (\gamma^{-1}\dot \gamma e_i).  \nonumber
\ee
The proposed action \cite{MPV18} is the following:
\begin{equation}
{S}= \int_R {L} dt=  \frac{1}{2}\int_{\mathbb{R}}\bigl( k_1\Braket{\gamma^{-1}\mathrm{d}\gamma\stackrel{\wedge}{,}* \gamma^{-1}\mathrm{d}\gamma} +k_2  ((\gamma^{-1}\mathrm{d}\gamma \stackrel{\wedge}{,} * \gamma^{-1}\mathrm{d} 
\gamma)) \bigr),
\label{newac}
\end{equation}
where $k_1,k_2$  are  real parameters,  and the two non-degenerate  scalar products in   Eqs. \eqn{Lprod},  \eqn{Rprod}, are employed, yielding 
\beqa
 \Braket{\gamma^{-1}\mathrm{d}\gamma \stackrel{\wedge}{,} * \gamma^{-1}\mathrm{d}\gamma}
 &=&     \dot {\bf Q}^I  \dot {\bf Q}^J  \eta_{IJ}\\
((\gamma^{-1}\mathrm{d}\gamma\stackrel{\wedge}{,} *\gamma^{-1}\mathrm{d}\gamma))&=& \dot {\bf Q}^I  \dot {\bf Q}^J \mathcal{H}_{IJ}
\label{prodotto2}
\eeqa
so that the generalized Lagrangian reads
\be \label{explito}
{L}=       \frac{1}{2}  ( k\, {\cal \eta}_{IJ} +  {\cal H}_{IJ})\dot {\bf Q}^I  \dot {\bf Q}^J  
 \ee
 with 
 \be
k\, {\cal \eta}_{IJ}+  {\cal H}_{IJ}=  
\begin{pmatrix}
\delta_{ij}& k \delta_i^j +    {\epsilon_{3i}}^{j} \\
-{\epsilon^i}_{j3}+k \delta_i^j& \delta^{ij}+ {\epsilon^i}_{l3} \epsilon^j_{k3}\delta^{lk}
\end{pmatrix}   \nonumber
\ee  
and where the  position $k_1/k_2 \equiv  k $ has been made. 
Proceeding as before we get the equations of motion in the form:
\be
{\sf L}_\Gamma\dot {\bf Q}^I( k\, {\cal \eta}_{IJ}+ \, {\cal H}_{IJ}) -\dot {\bf Q}^P \dot {\bf Q}^Q C_{IP}^K ( k\, {\cal \eta}_{QK}+ \,{\cal H}_{QK})  =0 \label{eomd}
\ee
where $C_{IP}^K$ are the structure constants of $\mathfrak{sl}(2,\C)$. The   matrix   $ k\, {\cal \eta}_{IJ}+ \, {\cal H}_{IJ}$ is non-singular, provided $k^2 \ne 1$, which will be assumed from now on. 

In \cite{MPV18} it has been shown that the Lagrangian of the IRR and of its dual model can be recovered by exploring the global symmetries of the generalized  dynamics.  If we choose  a local parametrization  for the elements of the double group $SL(2,\mathbb{C})$: $\gamma= \tilde{g}g$, with $g \in SU(2)$ and $\tilde{g} \in SB(2,\mathbb{C})$, 
  from Eq. \eqref{newac} it is easily seen  that $  L$ is invariant under left and  right action of the group $SU(2)$, 
 and  under left action of the group $SB(2,\mathbb{C})$. In order to recover the IRR Lagrangian we therefore gauge the  $SB(2,\C)_L$ global symmetry, by introducing a $SB(2,\C)$ gauge connection $\tilde C= \tilde {C_i}(t)\tilde e^i $: 
\be\label{sbgauge}
\gamma^{-1}d\gamma\rightarrow \gamma^{-1}D_{\tilde C}\gamma= (\gamma^{-1}\dot\gamma+\gamma^{-1}\tilde C\gamma) dt   
\ee
and performing the substitution
\be
\gamma^{-1}\dot\gamma+\gamma^{-1}\tilde C\gamma=\gamma^{-1}\dot\gamma+ \tilde{C_i} \gamma^{-1} \tilde e^i \gamma= \mathcal{U}_i\tilde e^i + \mathcal{W}^i e_i   \nonumber \,\,\,
\ee
from which the new variables $\mathcal{U}_i,  \mathcal{W}^i$ are easily retrieved. 
Analogously,  in order to obtain the Lagrangian of the dual model we gauge the global $SU(2)_R$ invariance, by introducing the $SU(2)$ gauge connection $C= C^i (t)  e_i$ so to have 
\be
\gamma^{-1}d\gamma\rightarrow \gamma^{-1}D\gamma=\tilde {\mathcal U}_i {\tilde e}^i + \tilde {\mathcal W}^i e_i .
\ee
By considering each of the  gauged Lagrangian functions, $L_{\tilde C}$ or $L_{C}$ and re-expressing them  in terms of the new variables, 
the partition function $Z$  is considered for each of them 
\begin{equation}
Z_1=\int \mathcal{D}g \mathcal{D}\tilde{g}\mathcal{D} {\tilde C} e^{-S_{{\tilde C}}} 
\label{partitt}
\end{equation}
or 
\begin{equation}
Z_2=\int \mathcal{D}g \mathcal{D}\tilde{g}\mathcal{D} { C} e^{-S_{{C}}} 
\label{partitt}
\end{equation}
and the  integration over the gauge potential $\tilde C$, respectively $C$, is performed.  Using techniques which are standard in field theory,  the integration with respect to the gauge potentials is traded for  the integration with respect to ${\mathcal U}_i$, $\tilde {\mathcal{W}}^i $ respectively, so that we are left with half the degrees of freedom of the generalized action \eqn{newac} and we retrieve the IRR model or the dual model, depending on which gauged Lagrangian we started with. We refer for details to \cite{MPV18}. 

A generalized kinematics in the context of DFT has been considered in \cite{freidel}.

\subsection{The Hamiltonian Formalism}\label{hamform} 
In the doubled description introduced above, the left generalized momenta are represented by:
\be
{\bf P}_I = \frac{\del  L}{\del \dot {\bf Q}^I}= ( {\cal \eta}_{IJ}+ k\,{\cal H}_{IJ})\dot {\bf Q}^J  \label{genP}
\ee
The Hamiltonian reads then as:
\be
{H}= ({\bf P}_I \dot{\bf Q}^I -  L)_{{\bf P}}= \frac{1}{2} [( {\cal \eta}+ k\, {\cal H})^{-1}]^{IJ} {\bf P}_I{\bf P}_J   \nonumber
\ee
with 
\be
[( {\cal \eta} + k\, {\cal H})^{-1}]^{IJ}= \frac{1}{2} (1-k^2)^{-1} 
\begin{pmatrix}
\delta^{ij}+ \epsilon^i_{l3} \epsilon^j_{k3}\delta^{lk}& -{\epsilon^i}_{j3}-k \delta^i_j
\\
{\epsilon_i}^{j3}-k \delta_i^j& \delta_{ij} \,\,\, 
\end{pmatrix}  \,\,.    \nonumber
\ee
From  \eqn{genP} one can explicitly write the generalized momenta ${\bf P}_I$ in terms of the components of $\dot{\bf Q}^I\equiv(A^i, B_j)$, finding:
\be
{\bf P}_I \equiv ( I_i, \tI^i)=\left(\delta_{ij} A^j+(k\delta_i^j+ \epsilon_i^{j3})B_j, (k \delta^i_j-\epsilon^i_{j3})A^j+[\delta^{ij}+\delta^{lk}\epsilon^i_{l3}\epsilon^j_{k3}]B_j\right).  \nonumber
\ee
In terms of the components $I_i, \tI^j$, it turns out that:
\be
{H} 
= 
\frac{1}{2}(1-k^2)^{-1} \left ( (1-k^2) \delta^{ij} I_i I_j + \delta_{ij}(\tI^i -I_s(k \delta^{si}+ {\epsilon^{si}}_3))
(\tI^j -I_r(k  \delta^{rj}+ {\epsilon^{rj}}_3))\right)\nn
\ee
In order to obtain the Hamilton equations for the generalized model on the Drinfel'd double,  one can proceed as in the previous sections with the determination of Poisson brackets from the first-order action functional \cite{MPV18}, which yields:
\beqa \label{remark}
\{I_i, I_j\}&=& {\epsilon_{ij}}^k I_k \\
\{\tI^i, \tI^j\}&=& {f^{ij}}_k \tI^k\\
\{I_i, \tI^j\}&=& {\epsilon^j}_{il} \tI^l- I_l {f^{lj}}_i  \;\;\;\;\{\tI^i, I_j\}= -{\epsilon^i}_{jl} \tI^l+ I_l {f^{li}}_j \label{remark3}
\eeqa
while the Poisson brackets between momenta and configuration space variables $g,\tg$ are unchanged with respect to $T^*SU(2), T^*SB(2,\C)$. We refer to \cite{MPV18} for details. 

Poisson brackets may be written in compact form:
\be
\{{\bf P}_I, {\bf P}_J\}= C_{IJ}^K {\bf P}_K   \nonumber
\ee
with $C_{IJ}^K$ the structure constants specified above. Thus we have, for Hamilton equations of motion: 
\be
\frac{d}{dt} {\bf P}_I= \{ {\bf P}_I, \widehat H\}=  [( {\cal \eta}+ k\, {\cal H})^{-1}]^{JK} \{ {\bf P}_I, {\bf P}_J\} {\bf P}_K= [( {\cal \eta}+ k\, {\cal H})^{-1}]^{JK} C_{IJ}^L {\bf P}_L{\bf P}_K   \nonumber
\ee
which is not zero, consistently with \eqn{eomd}.
 
\subsection{The Poisson Algebra}\label{canform}
The generalized formulation of the isotropic rotator is completed by discussing  the Poisson brackets on the double group $SL(2,\C)$, which correctly generalize those on the cotangent bundle stated in Eqs.  \eqn{pp}-\eqn{ij} as well as in Eqs. \eqn{ppdu}-\eqn{ijdu}. These have been introduced long time ago in \cite{semenov, alex} in the form
\be
\{\gamma_1,\gamma_2\}= -\gamma_1\gamma_2 r^* -r \gamma_1\gamma_2 \label{gammagamma2}
\ee
where $\gamma_1= \gamma\otimes 1, \gamma_2= 1\otimes \gamma_2$ while $r \in \mathfrak{d} \otimes \mathfrak {d}$ is the classical Yang-Baxter matrix:
 \be
r =  e^i\otimes e_i \label{rmatrix}
\ee 
satisfying the modified Yang-Baxter equation 
\be
[r_{12},r_{13}+r_{23}]+ [r_{13},r_{23}]= h   \nonumber
\ee
with  $r_{12 }=e^i\otimes e_i  \otimes \mathds{1}$, $r_{13}=e^i\otimes\mathds{1}\otimes  e_i $, $r_{23}= \mathds{1\otimes e^i\otimes e_i }$,  and  $h\in \mathfrak{d}\otimes \mathfrak{d}\otimes\mathfrak{d}$ and adjoint invariant element in the enveloping algebra.  The matrix 
\be
r^*= - e_i\otimes e^i \label{rmatrix2}
\ee
is also solution of the Yang-Baxter equation. 
On writing $\gamma$ as $\gamma= \tilde g g$ it can be shown that \eqn{gammagamma2} is compatible with the following choice 
\begin{align} 
 \{g_1,g_2\}  &= [r^*,g_1g_2], 
\label{pbm1}\\
\{{\tilde g}_1,g_2\} &=- {\tilde g}_1r  g_2 \label{finalpoi}\\
 \{\tilde g_1,\tilde g_2\}  &=-[r,\tilde g_1\tilde g_2],
\label{pbm2} 
\end{align}
with $g_1=g\otimes \mathds{1}$, $g_2=\mathds{1}\otimes g$, $\tilde g_1={\tilde g}\otimes \mathds{1}$ and ${\tilde g}_2= \mathds{1} \otimes {\tilde g}$.  Eqs. \eqn{pbm2} \eqn{pbm1} are the so-called Sklyanin brackets \cite{skly}. We also have $\{ { g}_1,{\tilde g}_2\} =- {\tilde g}_2 r^*   g_1$, with \eqn{pbm1}, \eqn{pbm2} Poisson-Lie brackets on $G$, $G^*$ respectively. Let us sketch how to recover  Eqs. \eqn{pp}-\eqn{ij} when passing from the double group $SL(2,\C)$ to either of the cotangent bundles. In order to recover the Poisson algebra for $T^*SU(2)$, one has to  rescale the matrix $r$  and the group elements of $SB(2,\C)$ by a real parameter  $\lambda$. By expanding up to first order, $ 
\tilde g(\lambda)=e^{i\lambda I_i e^i}   = 1+i\lambda I_i e^i + \mathcal{O}(\lambda^2) 
$ and replacing into  \eqref{pbm2}  we obtain:
\begin{equation}
\{\tilde g_1,\tilde g_2\}=\simeq -\lambda^2 e^i\otimes e^j \{I_i,I_j\} + \mathcal{O}(\lambda^3),   \nonumber
\end{equation}
which has to be equated to 
\be 
[r,\tilde g_1 \tilde g_2]\simeq  - \lambda^2 I_k  \epsilon^k_{\,ij}e_i\otimes e_j+ \mathcal{O}(\lambda^3)
\ee
thus yielding 
\begin{equation}
\{I_i,I_j\}=\epsilon^k_{\,ij}I_k.  \label{first}
\end{equation}
As for   Eq. \eqn{finalpoi} in order to compute its l.h.s. we use   the parametrization $g= y^0 \sigma_0 + i y^i \sigma_i$, so that,  up to first order in $\lambda$
\be
\{\tilde g_1,g_2\}=2 i\lambda\left( \{I_i, y^0\} e^i\otimes e_0+i \{I_i, y^j\} e^i\otimes e_j\right)+ O(\lambda^2)\label{lhs2}
\ee
while for the r.h.s.
\be
-\tilde g_1 r g_2\simeq - \lambda(  y^0 e^k\otimes e_k+ i y^j e^k\otimes (\delta_{kj}e_0+ i \epsilon_{kj}^i e_i)\label{rhs2}
\ee
thus yielding
\beqa\label{second}
\{I_i, y^0\}&=& - y^j \delta_{ij}  \nonumber \\
\{I_i, y^j\}&=&  y^0 \delta_i^j - y^k \epsilon_{ki}^j   \nonumber
\eeqa
where the first one is compatible with the second one, by using $(y^0)^2= 1- \sum_k y^k y^k$. 
Finally, on considering \eqref{pbm1} we observe that the LHS doesn't depend on $\lambda$ whereas the RHS does. Therefore we get
\be
\{y^0, y^j\} = \{y^i, y^j\} = 0 + O(\lambda) \,\,\, . \label{third}
\ee
Thus, Eqs. \eqn{first}, \eqn{second}, \eqn{third} reproduce correctly the canonical Poisson brackets on the cotangent bundle i$T^*SU(2)$. 

The Poisson brackets for the cotangent bundle $T^*SB(2,\C)$ are obtained in complete analogy, when considering  $r^*$ as an  independent solution of the Yang-Baxter equation  
\be
\rho= -\mu e_k\otimes e^k \label{altrmat}
\ee
and expanding $g\in SU(2)$ as a function of the parameter $\mu$. 
\be
g= \mathds{1} + i \mu \tI^i e_i + O(\mu^2)   \label{gexp} \,\,\, .
\ee

 \subsection{Poisson-Lie simmetries}\label{PLsym}
 Let us  explicitly address the nature  of symmetries  of the dual models introduced in the previous sections. In particular we want to discuss to what extent the models possess Poisson-Lie symmetries. We closely follow  \cite{marmo:articolo1}  for this subsection.   Poisson-Lie symmetries are Lie group transformations implemented on the carrier space of the dynamics  via group multiplication, which, in general,  are not canonical transformations as they need not preserve
the symplectic structure. However, if the Poisson structure is of the form \eqn{gammagamma2} with  carrier space  $D$ itself,  or \eqn{pbm1}, \eqn{pbm2} if we are looking at  $G$, $G^*$ respectively, Poisson brackets   can be made invariant if  the parameters of the group of transformations are imposed  to have
 nonzero Poisson brackets with themselves. Group multiplication is then said to
correspond to a Poisson map.  We have for example,  for the
right transformations of $G$ on
$D$,
\be\label{rightGact}
\gamma\rightarrow \gamma h \;,\; h \in G\;,\; \gamma \in D
\ee
and the left action of $G^*$ on $D$,
\be\label{leftG*act}
\gamma\rightarrow \tilde h \gamma  \;,\; h^* \in G^*\; \; \gamma \in D
.
\ee
In terms of the coordinates $(\tilde g,g)$ this implies
\be
g\rightarrow gh\;, \quad \tilde g\rightarrow \tilde g \;,
\ee
for the former and
\be
g\rightarrow g\;, \quad \tilde g\rightarrow \tilde h \tilde g  \;,
\ee
for the latter.  By themselves these
transformations  do not preserve the Poisson brackets
\eqn{pbm1}-\eqn{pbm2}.  But they can be made to be invariant
 if we require that the parameters of the transformation, $h$, have the following Poisson brackets
\be\label{hh}
\{h_1,h_2\}=[\;r^*\;,\;h_1 h_2\;] \;,
\ee
and zero Poisson brackets with $g$ and $\tilde g$.
Then $SU(2)$ right multiplication is a Poisson map and \eqn{rightGact} corresponds
to a Poisson Lie group transformation.
For \eqn{leftG*act} to be a Poisson Lie group transformation, $\tilde h$
must have the following Poisson bracket with itself
\be\label{hsthst}
\{\tilde h_1,\tilde h_2\}=-[\;r\;,\;\tilde h_1 \tilde h_2\;] \;,
\ee
and zero Poisson brackets with $g$ and $\tilde g$.
Since the right-hand-sides of \eqn{hh} and \eqn{hsthst} vanish
in the limit $\lambda \rightarrow 0 $,
the transformations \eqn{rightGact} and \eqn{leftG*act} become canonical in the limit.
\\
Moreover,  Poisson brackets \eqn{pbm1}-\eqn{pbm2}  are invariant under the
simultaneous action of both $G$ and $G^*$ via \eqn{rightGact} and \eqn{leftG*act}, if 
 we assume that{}\be
\{\tilde h_1,h_2\}=0\;.
\ee
By comparing with eq. \eqn{finalpoi} we
conclude that the algebra of the observables
$g$ and $\tilde g$ is different from the algebra of the symmetries
parametrized by $h$ and $\tilde h$.
Therefore, dynamics on the group manifold of $SL(2,\C)$ and on the two partner groups $SU(2)$ and $SB(2,\C)$ possesses  Poisson-Lie group symmetries, when endowed with the above mentioned brackets.  

Let us go back to the symplectic structures of  the IRR and the dual model, respectively given by Eqs. \eqn{xx} and \eqn{xxdu}. The former is obtained  from \eqn{pbm2} while the latter is obtained from  \eqn{pbm1}, for small (but non-zero) value of the parameters $\lambda$ and $\mu$, as we have shown in \ref{canform}. We can therefore conclude that the momentum variables of each model inherit their Poisson brackets from the Poisson-Lie structure  of the dual group, which  in turn exhibits  Poisson-Lie symmetry in the sense elucidated above.

\section{Conclusions}\label{concl}
Inspired by  an existing description of the dynamics of the Isotropic Rigid Rotator on a double Lie group  \cite{marmo:articolo1}, we have introduced a new dynamical model which is dual to the standard IRR. To this, we have used the notion of Poisson-Lie groups and Drinfel'd double for understanding the duality between the carrier spaces of the two models. Specifically, we have used the Drinfel'd double of the group $SU(2)$ as the  configuration space for the dynamics of a generalized model, with doubled degrees of freedom. This model exhibits non-Abelian duality and is an ideal arena to analyze in a simple context the meaning to physics of generalized and doubled geometry structures. 
Moreover, we have shown that, from the generalized action, the usual description with half the degrees of freedom, can be recovered by gauging one of its symmetries. 

The simple model of the IRR is especially interesting as a toy model for field theories with non-trivial target spaces such as Principal Chiral Models. In their original formulation \cite{gursey} these are nonlinear sigma models with the principal homogeneous space of the Lie group $SU(N)$ as its target manifold, where $N$ is the number of quark flavors.  An action functional, which is formally  analogous to the one introduced for the IRR can be written in the form 
\be
S=\frac{1}{2}\int_{\mathbb{R}^2}Tr(g^{-1}\mathrm{d}g \wedge * g^{-1}\mathrm{d}g),\label{chimoac}
\ee
where trace is understood as the scalar product on the Lie algebra $\mathfrak{g}.$  The Hodge operator exchanges the time and space derivatives
\be 
* (g^{-1}\mathrm{d}g)= *( \dot Q^i  dt+{Q^i}' d\sigma ) e_i= ( \dot Q^i  d\sigma-{Q^i}'dt ) e_i
\ee
with $ \dot Q^i =\Tr g^{-1}\partial_{t}g  e_i$, $ { Q^i}' =\Tr g^{-1}\partial_{\sigma}g  e_i$. It describes the dynamics of two dimensional fields $g: \mathbb{R}^{1,1}\rightarrow G$, with 
\be
g^{-1}\mathrm{d}g=g^{-1}\partial_{t}g \mathrm{d}t+g^{-1}\partial_{\sigma}g \mathrm{d}\sigma
\ee 
the Lie algebra valued left-invariant one-forms. 
Therefore, the dynamical fields of the model  take value in the cotangent bundle of the Lie group, while the canonical formalism is described by a Poisson algebra which takes the form of a semi-direct sum. The analogy with the IRR is thus very strict:  the analysis we have performed can be readily generalized, starting from an alternative description of Principal Chiral Models given in refs \cite{rajeev:bosonization, vitale1, vitale2, delduc} (also see \cite{reid, copland} where sigma  models are analyzed in the DFT context).  We are completing the analysis and  the results will be detailed in a forthcoming paper \cite{MPV18_2}. 

\noindent{\bf Acknowledgements} 
 P. V.  acknowledges  support by COST (European Cooperation in Science  and  Technology)  in  the  framework  of  COST  Action  MP1405  QSPACE.

\end{document}